\documentstyle[12pt]{article}
\title{\bf
             CURIE TEMPERATURES FOR THREE-DIMENSIONAL \\
             BINARY ISING FERROMAGNETS}
\author{
{\bf   Z. N\'eda } \\
{\small\it
      Babes-B\'olyai University, Dept. of Physics}   \\
{\small\it
      str. Kogalniceanu 1, RO-3400 Cluj, Romania  }   \\
     \and
{\small\it and  }\\
{\small\it
       University of Bergen, Dept. of Physics   }\\
{\small\it
         All\'egaten 55, N-5007 Bergen, Norway}  }
\date{}

\begin{document}

\maketitle

\begin{center}

Abstract\\
\end{center}

Using the Swendsen and Wang algorithm, high accuracy Monte Carlo simulations
were performed to study the concentration dependence of the Curie temperature
in binary, ferromagnetic Ising systems on the simple-cubic lattice. Our
results are in good agreement with known mean-field like approaches. Based
on former theoretical formulas we propose a new way of estimating the Curie
temperature of these systems.

PACS number(s): 75.10H; 75.20E; 75.40M

\vskip 40pt

\vfill \eject

\section*{I. Introduction}

Binary Ising systems presented a large interest from both bond and site
perspectives [1-6]. In the bond-disordered models the lattice sites
are equivalent
and the interaction energies between neighbouring sites are randomly assigned
from a set of possible values. In the site-disordered model the lattice
sites are randomly occupied by two different type of magnetic ions, $A$ and
$B$, with spins $S_A$ and $S_B$, and the interaction parameters between two
neighbouring spins are completly determined by their species. The randomness
in these systems can be considered either quenched or annealed. The annealed
systems proved to be handled theoretically much easier by mean-field like
methods, and so there are much better understood than the quenched ones.
Unfortunately for practical applications the quenched systems are much more
appropriate. This is the main reason why we proposed to limit our discussion
just for the case of quenched systems.

In the case of only ferromagnetic interactions between the spins, these models
were used with succes to describe the magnetic properties of quenched and
disordered magnetic alloys of the form $A_xB_{1-x}$, where $A$ and $B$ are
magnetic atoms \cite{6,7}. When antiferromagnetic and ferromagnetic
interactions
compete, frustration appears, and the system become a Mattis-Luttinger type
spin-glass model \cite{8,9}.

For real physical cases the site-disordered models are more realistic, and so
we proposed to study the Ising version of this model, considering the simplest
case of $S_A=S_B=\frac{1}{2}$, and all the exchange interactions of
ferromagnetic type.
The hamiltonian of our problem will be:
\begin{equation}
H=-\sum_{<i,j>} [J_{AA} \cdot \delta_{iA} \cdot \delta_{jA} + J_{BB} \cdot
\delta_{iB} \cdot \delta_{jB} + J_{AB}\cdot (\delta_{iA} \cdot \delta_{jB} +
\delta_{iB} \cdot \delta_{jA})]\cdot S_i^z \cdot S_j^z ,
\end{equation}
where $\delta_{ix}=1$ if the spin $i$ is of type $x$, and $0$ otherwise, and
the sum is refering to all nearest-neighbours. In this paper we consider the
real three-dimensional version of the model, and for results concerning the
two-dimensional case we propose \cite{10}.

The model considered by us was
already investigated by many authors, using different methods. The first
molecular-field approximations were due to Vonsovskii \cite{11,12}. Frustrated
systems were studied by Aharony using renormalization-group technics \cite{13}
and by Tatsumi with Monte Carlo simulations \cite{9}. The case of only
ferromagnetic interactions was studied using a mean-field like approach
by Kouvel \cite{14}, and with the coherent potential approximation by Foo and
Wu \cite{15}. Mean-field theoretical approaches were also made in the works of
Thorpe and McGurn \cite{3} respective Tahir-Kheli and Kawasaki \cite{2}.
Ishikawa and Oguchi \cite{4} considered a Bethe-Peierls approach
and in the work of Honmura, Khater, Fittipaldi and Kaneyoshi \cite{5} we find
an effective-field theory for the two-dimensional model. Monte Carlo
simulations
were performed by Scholten \cite{16} to study the critical temperatures of
two-dimensional, binary Ising ferromagnets in function of the
relative species concentration and the relative interaction energy between
unlike ions. Scholten also studied the phase diagram for the three-dimensional
problem on cubic lattices for frustrated systems \cite{17}, and included
next-nearest-neighbour interactions too. The phase diagrams of binary Ising
ferromagnets were studied by Thorpe and McGurn \cite{3} both in the
site-disorder and bond-disorder cases. They pointed out that the phase
diagrams can be usefully cataloged in terms of the initial slope
$\frac{\partial \ln T_c}{\partial q}$ of the transition temperature $T_c$
considered in function of concentration $q$, at the two points $q=0$ and $q=1$.
With the help of perturbation theory they also determined
the initial slopes for two-dimensional systems.
The phase diagrams of binary Ising systems with randomly
distributed exchange parameters were investigated by Kaneyoshi and Li using
effective-field theory with correlations \cite{18}. In the book from Vonsovskii
\cite{7} and in the paper from Luborsky \cite{6} one can find promising
comparisions between experimental data and mean-field type predictions.
Diluted systems, where one of the two components are non-magnetic, also
presented a field of interest [19-21]. Recently there has been much
interest in systems of mixed $S_A$ and $S_B$ spins, where $S_A \neq S_B$
[22-25].

In spite of all these earlier works there remained some not
complitly clarified questions even for the simplest ferromagnetic case. The
main problems are concerning the values of the critical exponents and the
determination of the critical temperature of the system in general cases. Our
work is intended to study the dependence of the critical temperature in
function of the system composition and  values of the coupling constants.
We do this in a review context by comparing our high-accuracy Monte Carlo
simulations with available theoretical formulas. In this manner we will give
a practically useful and easy method of approximating the Curie temperature
of these systems for general composition and general interaction parameters.
We will also check the validity and limitations of different mean-field type
approximations available for the Curie temperature of binary magnetic alloys.

\section*{II. Used theoretical formulas}

The localized model of ferromgnetism involving nearest-neighbour exchange
integrals has an attractive simplicity for describing some magnetic systems.
Although this approach for the magnetism in metallic systems is not completly
acceptable due to the partially itinerant nature of the magnetic electrons,
the obtained results are usually in good agreement with experimental data. In
the case of binary magnetic alloys we are in a similar situation. The localized
model based on the Heisenberg or Ising hamiltonian (1) with nearest-neighbour
exchange, or the molecular-field theories proved to be applicable in
describing the variation of the critical temperature in function of the alloys
composition.

The first formula based on the molecular-field approximation
was derived, as we stated earlier by Vonsovskii \cite{11,12}, and used with
success to describe transition temperatures of binary magnetic alloys. The
proposed formula was:
\begin{equation}
T_c(q)=T_c(A,A)-2 \cdot [T_c(A,A)-T_c(A,B)]\cdot q + [T_c(A,A)+T_c(B,B)-
2 \cdot T_c(A,B)] \cdot q^2 ,
\end{equation}
where $T_c(A,A)$ and $T_c(B,B)$ are the Curie temperatures of the pure $A$ and
$B$ systems, $T_c(A,B)$ is the Curie temperature for a pure system caracterized
with all exchange interactions  equal with the ones between the $A$ and $B$
magnetic ions ($J_{AB}$), $T_c(q)$ is the Curie temperature of the mixture,
and $q$ is the concentration of the $B$ component.

We mention here that the critical temperature $T_c$ for an Ising system
on the simple-cubic lattice, caracterized with $J$ exchange interaction
constants (considering just nearest-neighbour interactions) is given by
$T_c\approx 4.44425 \cdot \frac{J}{k_B}$, with $k_B$ the Boltzmann constant.

Using a phenomenological model based on mean-field theory suitably modified,
so that the individual atomic moments are allowed to vary in magnitude with
their local environment, and considering only nearest-neighbour interactions
Kouvel \cite{14} proposed the formula:
\begin{eqnarray}
 & T_c(q)=\frac{1}{2} \cdot [T_c(A,A) \cdot (1-q)+T_c(B,B) \cdot q] +
\nonumber \\
 & + \{ \frac{1}{4}\cdot
[T_c(A,A) \cdot  (1-q) - T_c(B,B) \cdot q]^2+ T_c(A,B)^2 \cdot q \cdot
(1-q) \} ^{\frac{1}{2}} .
\end{eqnarray}

In the work of Foo and Wu \cite{15} the disordered composition dependent
exchange interaction is treated in a coherent potential approximation (CPA).
In the limit of weak scattering their method give the mean-field like results,
but in the strong scattering limit they predict such effects as critical
concentration for the appearance of ferromagnetism in the diluted models
\cite{21}, which is not obtained in mean-field theories. They proposed the
following cubic equation for $T_c(q)$
\begin{eqnarray}
 & \alpha^2 \cdot T_c(q)^3 + \nonumber \\
 & +[\alpha \cdot (T_c(A,A)+T_c(B,B)+T_c(A,B))- \alpha
\cdot (1+\alpha) \cdot <T_c>] \cdot T_c(q)^2+ \nonumber \\
 & + [(1+\alpha) \cdot T_c(A,A) \cdot
T_c(B,B) \cdot T_c(A,B) \cdot < \frac{1}{T_c} > - \nonumber \\
 & -\alpha \cdot (T_c(A,A) \cdot
T_c(B,B) + T_c(A,B) \cdot T_c(A,A) + T_c(A,B) \cdot T_c(B,B))] \cdot T_c(q) -
\nonumber \\
 & -T_c(A,A) \cdot T_c(B,B) \cdot T_c(A,B)=0,
\end{eqnarray}
where
\begin{equation}
\alpha=\frac{z}{2}-1,
\end{equation}
with $z$ the coordination number of the lattice (in our case $z=6$), and
\begin{eqnarray}
 & <T_c>=(1-q)^2 \cdot T_c(A,A) + 2 \cdot q \cdot (1-q) \cdot T_c(A,B) +
q^2 \cdot T_c(B,B) ,\\
 & <\frac{1}{T_c}>= \frac{(1-q)^2}{T_c(A,A)}+ \frac{2\cdot q \cdot
(1-q)}{T_c(A,B)}
+ \frac{q^2}{T_c(B,B)}.
\end{eqnarray}

We mention that there are also other, more evoluate possibilities of
calculating
the Curie temperature, based on the Ising model (1) of the system, such as
mean-field like renormalization-group technics, series expansion and
perturbation methods. Unfortunately these are all very technical ones, and do
not give practically usable formulas.

\section* {III. The computer simulation method}

As stated earlier, Monte Carlo simulations were performed on the considered
model (1) in the ferromagnetic case ($ T_c(A,A)>0,\: T_c(A,B)>0$ and
$T_c(B,B)>0$ ) by Scholten \cite{16}, and on frustrated systems by Tatsumi
\cite{9} and Scholten \cite{17}. Scholtens work for purely ferromagnetic
systems refers to the two-dimensional case. He used the classical single
spin-flip Metropolis algorithm \cite{26}, and due to this his calculations
were rather time-consuming. So, he considered just a few choices for the
interaction parameters. He compared his Monte Carlo results with the ones
obtained in \cite{2,3}, \cite{4} and \cite{5}.

In the present work we proposed to study by high-accuracy Monte Carlo
simulations the three-dimensional case of simple-cubic lattices, completing
in some sense the earlier works. We used the more performant cluster-flip
Swendsen and Wang Monte Carlo method \cite{27} with an original recursion type
algorithm. We  proposed to compare our results with the ones given in
\cite{11,12}, \cite{14} and \cite{15}.

Our simulations were performed on relatively large $50 \times 50 \times 50$
simple-cubic lattices. The critical temperature was found by detecting the
maximum in the fluctuation of the absolute value of the magnetization. For
achieving statistical equilibrium we considered up to $600$ cluster-flips
and then studied the fluctuation for $1000$ more iterations. The sensitivity
in the determination of the critical temperature was in general of the
order of $0.01 \cdot T_c(A,A)$. We chose usually $T_c(A,A)=100$ (units) and
we proposed to consider various values for the $T_c(A,B)$ and $T_c(B,B)$
parameters. For every chosen set of interaction parameters we covered the
$q \in (0,1)$ concentration interval uniformly with $19$ simulation points.
The program was written in C and the simulations were performed on a
CRAY Y-MP4D/464 computer and IBM R-6000 RISC workstations.

\section*{IV. Results}

Our Monte Carlo results for the variation of the Curie temperature in function
of the $B$ components concentration, are plotted with different symbols on
Fig. 1-7. The simulations, as we stated, earlier were made on a simple-cubic
lattice. The curves indicate theoretical results obtained from equations (2)
and
(3).

In Fig. 1 considering four choices of the $J_{AB}$ interaction parameters
($J_{BB}$ and $J_{AA}$ fixed) we compare our Monte Carlo results with the
ones obtained from equation (2).
Fig. 2 presents the same results in comparision with theoretical data given
by equation (3). One can observ that formula (2) predict in general lower
values, and in contrast to this (3) predict higher values for the transition
temperature than the real ones. We also checked, that equation (4) gives
lower values even than (2), so it is much less appropriate for our model.
As a first observation we conclude, that in these cases the real transition
temperatures are limited with the two curves given by equations (2) and (3).
We are also able to confirm, that in these three-dimensional cases the
mean-field like results proved to give quite good estimates for the Curie
temperature. In Fig. 3 we illustrate that almost a perfect fit with the
realistic Curie temperatures can be obtained, if we use the arithmetic mean
of the values obtained from equations (2) and (3).

In Fig. 4-6 we tried to prove our previous statements considering quite
exotic values for the exchange intercation parameters, and thus
for the $T_c(A,B)$ and $T_c(B,B)$ critical temperatures. We draw with
thin dashed lines the results given by equations (2) and (3) (dense dashes
correspond to the curve obtained from (2)). The continuous darker line
represents the arithmetic mean of formulas (2) and (3). We conclude again
that in general the values given by equations (2) and (3) limit nicely our
realistic simulation points, and their arithmetic mean give a fairy good
estimate for the realistic Curie temperature. In  the case when
$J_{AB} \not\in [J_{AA},\:J_{BB}]$,
one can also observ, that the strongest difference between the
arithmetic mean proposed by us and the
simulation results are for concentration values where the critical temperature
has an extremum, the real values being lower.

In Fig. 7 we present studies concerning the extrem case, when the $J_{AB}$
interaction parameter, and thus the $T_c(A,B)$ critical temperature is
getting smaller (weak coupling between the two components). In this case, as
expected, we get that the simulation curves in the limit of $q=0$ and $q=1$
are tending to strait-lines with the slope $\frac{1}{T_c(0)}\cdot \frac
{\partial T_c(q)}{\partial q}$, equal with $1.13$, characteristic for
site-diluted systems \cite{21}.

\section*{V. Conclusions}

 From comparision of computer simulation data with results given by formula
(2) and (3) we conclude, that in the three-dimensional case, the mean-field
like approaches are working satisfactory well. So, in this way it is not
surprising the good fit of the mean-field like predictions with experimental
data, presented in \cite{6} and \cite{7}.

Generally the curves obtained for the critical temperature from equation (2)
and (3) limit rather nicely the real values. Exceptions are the cases where
the $J_{AB}$ interaction parameter is not from the interval limited by $J_{AA}$
and $J_{BB}$. In these situations in the vicnity of the extremum of the
Curie temperature curve our previous statement might not be true.

The theoretical curve constructed from the arithmetic mean of formula (2) and
(3) proved to be a good approximation to obtain easily the Curie temperature
for quenched, binary Ising ferromagnets.

In the limit of small couplings beetween the two components ($J_{AB}$ small),
we obtained results in good agreement with the site-diluted model.

Our study is intended to complete the earlier ones by giving a practically
useful method of estimating the Curie temperature of the proposed system.
We also illustrated the validity of our method, and tried to study many
possible choices for the interaction parameters.

\section*{Acknowledgements}

\samepage

This study was finished during a bursary offered by the Norwegian Research
Council. We thank Y. Brechet, A. Coniglio,
L. Csernai, and L. Peliti for their continuous help and useful discussions.

\newpage
\section*{Figure Captions}
\vspace{.10in}

$\:$ \\
\vspace{.15in}

{\bf Fig. 1.} Monte Carlo results for the variation of the Curie temperature as
a function of the $B$ components concentration for four choices of the
$T_c(A,B)$ critical temperature. The values of $T_c(A,A)=100$ and
$T_c(B,B)=200$
are fixed. Solid curve is given by equation (2).
\vspace{.25in}

{\bf Fig. 2.}  The Monte Carlo results from  Fig. 1 in comparision with the
curve given by equation (3).
\vspace{.25in}

{\bf Fig. 3.}  The Monte Carlo results from Fig. 1 in comparision with the
Curie temperature given by the arithmetic mean of formulas (2) and (3).
\vspace{.25in}

{\bf Fig. 4.}  The dots and triangles represents Monte Carlo simulations
for the given values of the $T_c(A,B)$ critical temperature.
The thin dashed lines indicate the results obtained
from formulas (2) and (3) (dense dashes corresponding to (2)). The
continuous dark line indicate the Curie temperatures obtained as the arithmetic
mean of (2) and (3).
\vspace{.25in}

{\bf Fig. 5.}  The case when we have no excange interactions between the
atoms of the $B$ component ($J_{BB}=0$) and $T_c(A,A)=T_c(A,B)=100$. Dots
are Monte-Carlo results and the curves have the same meaning as in Fig. 4.
\vspace{.25in}

{\bf Fig. 6.}  Monte Carlo results (dots) for $T_c(A,A)=100$, $T_c(B,B)=500$
and $T_c(A,B)=50$. The curves represents the same formulas as in Fig. 4.
\vspace{.25in}

{\bf Fig. 7.} The upper figure shows Monte Carlo results for three small
choices of the $T_c(A,B)$ critical temperature.  Solid curves indicate the
characteristic strait-lines for site-diluted systems in the small dilution
limit.
The figure from below presents  Monte Carlo results obtained
for the site-diluted system.

\end{document}